\documentclass[letterpaper, 10 pt, conference]{ieeeconf}  

\IEEEoverridecommandlockouts                              

\usepackage{amssymb}
\usepackage{graphicx}
\usepackage{times} 
\usepackage{amsmath} 
\usepackage{cleveref}
\usepackage{multicol}
\usepackage{multirow}
\usepackage{cite}
\usepackage{balance}
\usepackage[affil-it]{authblk}
\usepackage{xcolor}
\crefformat{figure}{Fig.~#1#2}
\title{\LARGE \bf
ClusMFL: A Cluster-Enhanced Framework for Modality-Incomplete Multimodal Federated Learning in Brain Imaging Analysis
}

\author{Xinpeng Wang$^\dagger$, Rong Zhou$^\ddagger$, Han Xie$^\S$, Xiaoying Tang$^{\ast\dagger}$, Lifang He$^\ddagger$ and Carl Yang$^\S$
\thanks{*Xiaoying Tang is the corresponding author.}
\thanks{$^\dagger$Xinpeng Wang and Xiaoying Tang are with the School of Science and Engineering, the Chinese University of Hong Kong, Shenzhen. {\tt\small xinpengwang@link.cuhk.edu.cn}, {\tt\small tangxiaoying@cuhk.edu.cn}}%
\thanks{$^\ddagger$Rong Zhou and Lifang He are with the Department of Computer Science and Engineering, Lehigh University. {\tt\small \{roz322, lih319\}@lehigh.edu}}%
\thanks{$^\S$Han Xie and Carl Yang are with the Department of Computer Science, Emory University. {\tt\small\{han.xie, j.carlyang\}@emory.edu} 
}
}
\begin{document}

\maketitle
\thispagestyle{empty}
\pagestyle{empty}

\begin{abstract}

Multimodal Federated Learning (MFL) has emerged as a promising approach for collaboratively training multimodal models across distributed clients, particularly in healthcare domains. In the context of brain imaging analysis, modality incompleteness presents a significant challenge, where some institutions may lack specific imaging modalities (e.g., PET, MRI, or CT) due to privacy concerns, device limitations, or data availability issues. While existing work typically assumes modality completeness or oversimplifies missing-modality scenarios, we simulate a more realistic setting by considering both client-level and instance-level modality incompleteness in this study. 
Building on this realistic simulation, we propose ClusMFL, a novel MFL framework that leverages feature clustering for cross-institutional brain imaging analysis under modality incompleteness. Specifically, ClusMFL utilizes the FINCH algorithm to construct a pool of cluster centers for the feature embeddings of each modality-label pair, effectively capturing fine-grained data distributions. These cluster centers are then used for feature alignment within each modality through supervised contrastive learning, while also acting as proxies for missing modalities, allowing cross-modal knowledge transfer. Furthermore, ClusMFL employs a modality-aware aggregation strategy, further enhancing the model’s performance in scenarios with severe modality incompleteness. 
We evaluate the proposed framework on the ADNI dataset, utilizing structural MRI and PET scans. 
Extensive experimental results demonstrate that ClusMFL achieves state-of-the-art performance compared to various baseline methods across varying levels of modality incompleteness, providing a scalable solution for cross-institutional brain imaging analysis.
\end{abstract}

\section{INTRODUCTION}
Multimodal Federated Learning (MFL) has emerged as a transformative approach for collaboratively training machine learning models across distributed clients with multimodal data, especially in privacy-sensitive domains like healthcare \cite{MLF_healthcare_review,MFL_a_survey,survey_of_MFL}. By integrating diverse data modalities, MFL facilitates the development of robust and accurate models for intelligent clinical decision support systems. However, real-world healthcare applications, particularly in cross-institutional brain imaging analysis, often face the challenge of modality incompleteness \cite{FedMI}. Specifically, some institutions  may have access solely to PET imaging data, while others may have access only to MRI data. This disparity in available modalities significantly undermines the performance of traditional federated learning frameworks, which typically assume that all clients have access to the same set of modalities.

To address this challenge, researchers have proposed various methods to mitigate the impact of missing modalities in MFL. These methods can be broadly categorized into prototype-based approaches and generative approaches, each with distinct strengths and limitations. Prototype-based methods \cite{FedMI,MFCPL,PmcmFL} leverage the average of feature embeddings within the same class to serve as prototypes for feature alignment or modality completion. While this approach is effective in capturing general class-level characteristics, it struggles to encapsulate the inherent diversity and complexity of individual data points \cite{prototype_limiation1,prototype_limiation2,prototype_limiation3}. As a result, the simplistic averaging process often produces prototypes that inadequately represent the underlying data distribution, leading to suboptimal alignment of modality-specific feature embeddings and ultimately degrading the global model's performance.

On the other hand, generative approaches attempt to reconstruct missing modalities using generative models, which hold promise for improving data completeness \cite{FedMed-GAN,CACMRN, wu2024deep}. However, these generative methods typically require modality-complete instances for effective training, yet such instances are often scarce in real-world settings of modality-incomplete MFL \cite{CACMRN}. This limitation restricts their applicability to incomplete data and increases the likelihood of introducing noise or inaccuracies during the reconstruction process. Moreover, methods based on Generative Adversarial Networks (GANs) \cite{goodfellow2020generative} are particularly susceptible to mode collapse,  where the model fails to capture the full diversity of the data, leading to the generation of highly similar, non-representative samples \cite{model_collapse1,model_collapse2}. These limitations render generative approaches less robust and reliable in practical applications of modality-incomplete MFL.

Beyond the limitations of prototype-based and generative approaches in addressing modality incompleteness in MFL, existing studies on modality-incomplete MFL share several common shortcomings from the perspective of federated learning. One such limitation is the unrealistic simulation of MFL scenarios, which assumes uniform modality availability or consistent missing patterns across all clients. These oversimplified simulations fail to accurately represent real-world modality incompleteness scenarios, thereby limiting their practical applicability, particularly in cross-institutional brain imaging analysis \cite{FedMAC}. Furthermore, much of the current work on MFL with missing modalities employs uniform aggregation weights for different modules within the model during the federated learning process \cite{10654835}. This uniform approach overlooks the varying modality distributions across clients, resulting in suboptimal global model performance.



To more accurately reflect the modality incompleteness in real-world scenarios of cross-institutional brain imaging analysis, this paper introduces a realistic and comprehensive setting for modality distribution. Unlike prior studies that assume uniform modality availability or consistent missing modalities across clients, our setting simulates both client-level and instance-level modality incompleteness, as illustrated in \Cref{fig:setting}, capturing the diversity and complexity of modality distribution in real-world applications.

Building on this simulation, we propose ClusMFL, a novel framework designed to address the challenges of modality incompleteness in MFL. Concretely, we apply the FINCH clustering algorithm \cite{FINCH} to cluster feature embeddings and construct a pool of cluster centers for each pair of modality and label across clients, providing a finer-grained representation of data distributions compared to traditional prototype-based methods. To ensure that modality-specific encoders accurately extract modality-specific features related to the labels, we perform feature alignment by employing supervised contrastive loss \cite{supervised_contrastive_loss} over local feature embeddings and the pool of cluster centers. Furthermore, to mitigate the impact of severe modality incompleteness, we use these cluster centers as proxies for the feature embeddings of the missing modalities during the training process, enabling cross-modality knowledge transfer. In addition, we utilize a modality-aware aggregation method that assigns different aggregation weights to various modules of the model based on the modality distributions, effectively balancing the contributions from each client to different modules.

To evaluate the effectiveness of the proposed framework, we conduct extensive experiments under various settings of modality incompleteness and compare ClusMFL with several traditional federated learning algorithms, as well as algorithms specifically designed for modality-incomplete MFL. 
Specifically, we use brain imaging data from the real-world ADNI dataset, which includes structural MRI and PET scans for 915 participants stratified into three diagnostic categories: healthy controls, mild cognitive impairment (MCI), and Alzheimer’s disease (AD) patients. By simulating realistic settings of modality incompleteness on this dataset, we aim to evaluate our framework under conditions that reflect real-world clinical challenges. 
The results show that ClusMFL outperforms existing approaches, particularly in scenarios with substantial modality incompleteness, underscoring its capability to address real-world challenges in MFL.

\section{MODALITY INCOMPLETENESS SETTING}
\label{setting}
\begin{figure}[t] 
    \centering
    \includegraphics[width=\linewidth]{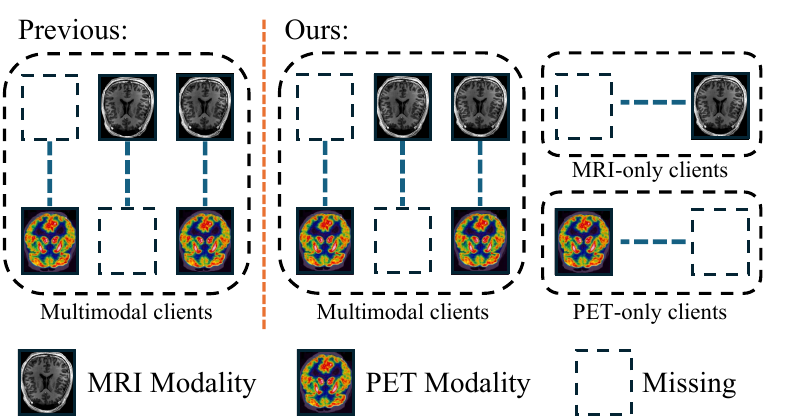} %
    \caption{Illustration of Modality Incompleteness Setting. Dash boxes denote the missing modality, while each pair of boxes represents an instance. }
    \label{fig:setting}
\end{figure}

In this study, we focus on two representative brain imaging modalities: PET and MRI, and simulate a more realistic setting by considering both client-level and instance-level modality incompleteness. Each instance is represented as a triplet \( (\mathbf{x}_P, \mathbf{x}_M, y) \), where \( \mathbf{x}_P \) corresponds to data from the PET modality, \( \mathbf{x}_M \) corresponds to data from the MRI modality, and \( y \) represents the associated label. Based on the availability of modalities, data instances are categorized into the following three types, as shown in \Cref{fig:setting}:

\begin{enumerate}
    \item \textbf{PET-only instances:} These instances contain data solely from the PET modality, with no corresponding MRI data. Such instances are formally denoted as \( d_P = (\mathbf{x}_P, \varnothing, y) \).

    \item \textbf{MRI-only instances:} These instances exclusively contain data from the MRI modality, with no accompanying PET data. These are represented as \( d_M = (\varnothing, \mathbf{x}_M, y) \).

    \item \textbf{Multimodal instances:} These instances include data from both modalities, and are represented as \( d_B = (\mathbf{x}_P, \mathbf{x}_M, y) \).
\end{enumerate}

At the client level, we categorize clients into three groups based on the composition of their instances:

\begin{enumerate}
    \item \textbf{PET-only clients:} These clients exclusively host PET-only instances. Their datasets are denoted as \( \mathcal{D}_P = \bigl\{d_P^i\bigr\}_{i=1}^{n} \), where \( n \) is the number of instances.

    \item \textbf{MRI-only clients:} These clients solely host MRI-only instances, denoted as \( \mathcal{D}_M = \bigl\{d_M^i\bigr\}_{i=1}^{n} \).

    \item \textbf{Multimodal clients:} These clients contain a mix of all three types of instances. Their datasets are represented as:
    \[
    \mathcal{D}_B = \bigl\{d_P^i\bigr\}_{i=1}^{\beta_1 n} \cup \bigl\{d_M^i\bigr\}_{i=1}^{\beta_2 n} \cup \bigl\{d_B^i\bigr\}_{i=1}^{(1-\beta_1-\beta_2)n},
    \]
    where \( \beta_1 \) and \( \beta_2 \) indicate the proportions of PET-only and MRI-only instances on multimodal clients, respectively, while \( 1-\beta_1-\beta_2 \) denotes the proportion of multimodal instances on multimodal clients.
\end{enumerate}

For client $i$, we denote the number of instances on this client as $n_i$. The overall dataset is defined as follows:

\begin{equation}
    \mathcal{D} = \bigcup_{i=1}^{\alpha_1 N} \mathcal{D}_P^i \cup \bigcup_{i=1}^{\alpha_2 N} \mathcal{D}_M^i \cup \bigcup_{i=1}^{(1-\alpha_1-\alpha_2) N} \mathcal{D}_B^i,
\end{equation}
where \( \alpha_1 \), \( \alpha_2 \), and \( 1-\alpha_1-\alpha_2 \) represent the proportions of PET-only, MRI-only, and multimodal clients, respectively, and \( N \) denotes the total number of clients participating in the federated learning framework.


\section{METHOD}
\begin{figure*}[htbp]
    \centering
    \includegraphics[width=\textwidth]{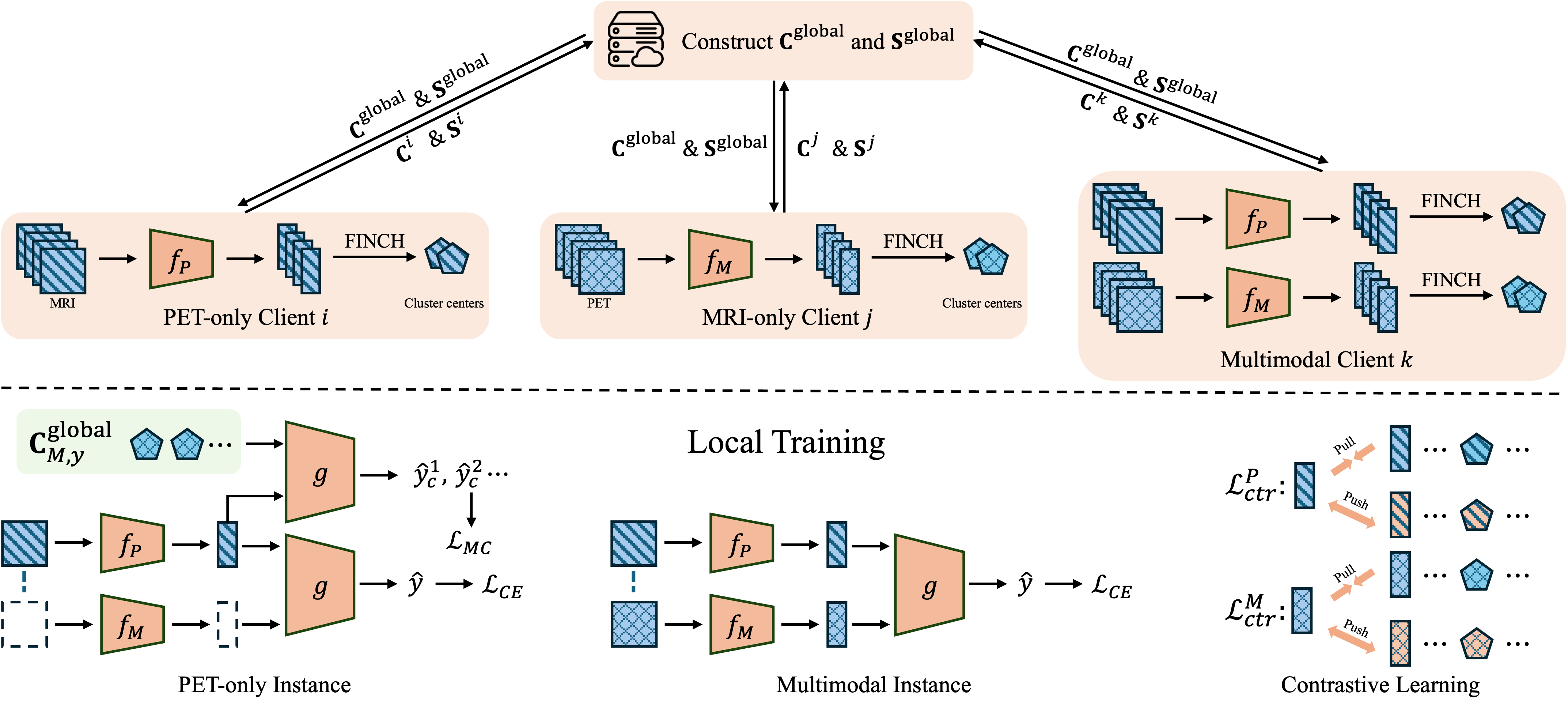} 
    \caption{Overview of ClusMFL. In this figure, PET-only instances are used as examples of single-modality instances in local training. Different patterns represent different modalities, and different colors indicate different labels.}  
    \label{fig:overview}
\end{figure*}
\subsection{Preliminary}
\label{sec: preliminary}
In this study, we adopt a typical architecture for multimodal models, which includes two encoders—one for each modality—and a classifier. Let \( f_P \) and \( f_M \) denote the encoders for PET modality and MRI modality, respectively. The encoders \( f_P \) and \( f_M \) are responsible for extracting the relevant features from each modality. The model also includes a classifier, denoted as \( g \), which concatenates the embeddings from the encoders and performs the final prediction.

During the inference stage, for instances containing both modalities, the input data \( \mathbf{x}_P \) and \( \mathbf{x}_M \) are processed through their respective encoders \( f_P \) and \( f_M \). Specifically, the feature embeddings are obtained as \( \mathbf{z}_P = f_P(\mathbf{x}_P) \) and \( \mathbf{z}_M = f_M(\mathbf{x}_M)\), which are subsequently concatenated and passed to the classifier \( g \) for the final prediction, \( \hat{y} = g(\mathbf{z}_P, \mathbf{z}_M) \). For instances with only a single modality, the feature embedding of the missing modality is replaced with a tensor of zeros, denoted as \( \mathbf{0} \). 

The federated training process repeats the construction of the pool of cluster centers and cluster sizes (\textit{i.e.}, \( \mathbf{C}^{\text{global}} \) and \( \mathbf{S}^{\text{global}} \)), followed by local training and aggregation in each round, which are explained in detail in the following sections. We provide an overview of the construction of  \( \mathbf{C}^{\text{global}} \) and \( \mathbf{S}^{\text{global}} \) and local training in \Cref{fig:overview}.

\subsection{Constructing The Pool of Cluster Centers}
Unlike traditional prototype learning methods, which use the mean of features as the prototype and result in a shift between individual samples and the prototype, this study adopts the FINCH \cite{FINCH} algorithm to construct a pool of cluster centers, thereby more effectively representing the clients' data distribution.

On each client, we begin by applying the FINCH clustering algorithm to calculate the local cluster centers for feature embeddings of each pair of modality and label. The resulting cluster centers, along with cluster sizes, are then uploaded to the server. For instance, consider the modality PET. The feature embeddings associated with modality PET are extracted as \( \mathbf{Z}_P = f_P(\mathbf{X}_P) \), where $\mathbf{X}_P$ represents all PET data on this client. The FINCH algorithm is then applied to \( \mathbf{Z}_P \) for clustering, yielding the corresponding cluster centers and cluster sizes. Specifically, for label \( j \) in modality PET, FINCH identifies the cluster centers as:
\begin{equation}
    (\mathbf{C}_{P,j},\mathbf{S}_{P,j})= \text{FINCH}(\mathbf{Z}_{P,j}),
\end{equation}
where \( \mathbf{Z}_{P,j} \) denotes the feature embeddings with modality PET corresponding to label \( j \), and \( \mathbf{C}_{P,j} = ( \mathbf{c}_{P,j}^{1}, \mathbf{c}_{P,j}^{2}, \dots, \mathbf{c}_{P,j}^{K_{P,j}} )\) represents the \( K_{P,j} \) cluster centers obtained from the FINCH algorithm for label \( j \). The set \( \mathbf{S}_{P,j} =( s_{P,j}^{1}, s_{P,j}^{2}, \dots, s_{P,j}^{K_{P,j}} ) \) represents the sizes of the corresponding clusters, where each \( s_{P,j}^{k} \) denotes the number of feature embeddings assigned to the \( k \)-th cluster for label \( j \), with \( \mathbf{c}_{P,j}^{k} \) being the cluster center. $K_{P,j}$ represents the number of clusters, which is determined automatically by FINCH.

For client \(i\), the cluster centers and sizes associated with modality PET and label \(j\) are denoted as \(\mathbf{C}_{P,j}^i\) and \(\mathbf{S}_{P,j}^i\), respectively. If client \(i\) lacks data for modality PET, we set \(\mathbf{C}_{P,j}^i = \varnothing\) and \(\mathbf{S}_{P,j}^i = \varnothing\).

Once the cluster centers and cluster sizes for each client are computed, they are sent to the server, where they are collected to form a global pool of cluster centers. Specifically, the global pool of cluster centers for label \( j \) in modality PET, denoted as \( \mathbf{C}_{P,j}^{\text{global}} \), is constructed by concatenating the cluster centers from all clients:

\begin{equation}
    \mathbf{C}_{P,j}^{\text{global}} = \bigoplus_{i=1}^N \mathbf{C}_{P,j}^i.
\end{equation}
Similarly, the corresponding cluster sizes for label \( j \) are cancatenated into a global pool, denoted as \( \mathbf{S}_{P,j}^{\text{global}} \):

\begin{equation}
    \mathbf{S}_{P,j}^{\text{global}} = \bigoplus_{i=1}^N \mathbf{S}_{P,j}^i.
\end{equation}

The global pools \( \mathbf{C}_{P,j}^{\text{global}} \) and \( \mathbf{S}_{P,j}^{\text{global}} \) encapsulate the aggregated cluster centers and their respective sizes for each label \( j \) across all clients, enabling the model to leverage a comprehensive representation of the data distribution. These global pools are then distributed back to each client, allowing them to utilize the global information during local training.

The same process is performed for MRI modality. Let \( \mathbf{C}_{M,j}^i \) and \( \mathbf{S}_{M,j}^i \) represent the cluster centers and cluster sizes, respectively, for MRI and label \( j \) on client \( i \). The global pools for MRI are constructed as:

\begin{equation}
    \mathbf{C}_{M,j}^{\text{global}} = \bigoplus_{i=1}^N \mathbf{C}_{M,j}^i, \quad \mathbf{S}_{M,j}^{\text{global}} = \bigoplus_{i=1}^N \mathbf{S}_{M,j}^i.
\end{equation}

These global pools for MRI are also distributed back to the clients to ensure that the global knowledge from both modalities is available for further training and optimization.

\subsection{Feature Alignment}

To ensure that the encoders \( f_M \) and \( f_P \) extract the correct modality-specific features and mitigate overfitting under severe modality incompleteness, we combine global cluster centers and local feature embeddings and apply supervised contrastive loss for feature alignment.

For any multimodal client with dataset $\mathcal{D}_B$, let \( \mathbf{X}_P \) and \( \mathbf{X}_M \) represent all PET and MRI data on this client, respectively, along with their corresponding labels \( \mathbf{y}_P \) and \( \mathbf{y}_M \). The feature embeddings for PET modality and MRI modality are computed as \( \mathbf{Z}_P = f_P(\mathbf{X}_P) \) and  \( \mathbf{Z}_M = f_M(\mathbf{X}_M) \) respectively. To incorporate global information, the feature embeddings \( \mathbf{Z}_P \) are concatenated with the global cluster centers \( \mathbf{C}_{P}^{\text{global}} \):
\begin{equation}
    \mathbf{Z}_{P,G} = \mathbf{Z}_P \oplus \bigoplus_{j=1}^J \mathbf{C}_{P,j}^{\text{global}},
\end{equation}
where \( J \) is the total number of unique labels. Similarly, the label \( \mathbf{y}_P \) is extended as:
\begin{equation}
    \mathbf{y}_{P,G} = \mathbf{y}_P \oplus \bigoplus_{j=1}^J (j )^{|\mathbf{C}_{P,j}^{\text{global}}|},
\end{equation}
where \( ( j )^{|\mathbf{C}_{P,j}^{\text{global}}|} \) denotes a list containing label \( j \) repeated \( |\mathbf{C}_{P,j}^{\text{global}}| \) times.

We compute the supervised contrastive loss over \( \mathbf{Z}_{P,G} \). Let \( \mathcal{I} = \{ 1, 2, \dots, |\mathbf{y}_{P,G}| \} \) denote the index set, and define \( \mathcal{A}(i) = \mathcal{I} \backslash \{ i \} \) as the set of all indices excluding \( i \). The supervised contrastive loss for PET modality is defined as:
\begin{equation}
    \mathcal{L}_{\text{ctr}}^P(\mathbf{X}_P) = \mathbb{E}_{i \in \mathcal{I}} \left[ -\frac{1}{|\mathcal{P}(i)|} \sum_{p \in \mathcal{P}(i)} \log \frac{v_{i,p}}{\sum_{a \in \mathcal{A}(i)} v_{i,a}} \right],
\end{equation}
where:
\begin{itemize}
    \item \( \mathcal{P}(i) = \{ p \in \mathcal{A}(i) \mid \mathbf{y}_{P,G}^{(p)} = \mathbf{y}_{P,G}^{(i)} \} \) represents the set of indices corresponding to positive examples for \( \mathbf{Z}_{P,G}^{(i)} \).
    \item \( v_{i,p} = \exp(\text{sim}(\mathbf{Z}_{P,G}^{(i)}, \mathbf{Z}_{P,G}^{(p)}) / \tau) \), where \( \text{sim}(\cdot, \cdot) \) denotes the cosine similarity between two embeddings, and \( \tau \) is a temperature parameter.
\end{itemize}

The similar process is applied to calculate \( \mathcal{L}_{\text{ctr}}^M(\mathbf{X}_M) \) and the overall supervised contrastive loss for $\mathcal{D}_B$ is then computed as:
\begin{equation}
    \mathcal{L}_{\text{CTR}} (\mathcal{D}_B)= 
    \frac{
        |\mathbf{y}_P| \cdot \mathcal{L}_{\text{ctr}}^P(\mathbf{X}_P) 
        + 
        |\mathbf{y}_M| \cdot \mathcal{L}_{\text{ctr}}^M(\mathbf{X}_M)
    }{
        |\mathbf{y}_P| + |\mathbf{y}_M|
    }.
\end{equation}

For clients with data from only one modality, the loss is computed solely for the corresponding modality, without involving the other modality. 

\subsection{Modality Completion Loss}
To further enhance the model's classification performance in the presence of missing modalities, we use $\mathbf{C}^{\mathrm{global}}$ as approximations for feature embeddings of the missing modality. Specifically, for a PET-only instance with data \( d_P = (\mathbf{x}_P, \varnothing, y) \) or an MRI-only instance with data \( d_M = (\varnothing, \mathbf{x}_M, y) \), we first pass the available modality through the corresponding encoder to obtain the feature embedding \( \mathbf{z}_P \) and \( \mathbf{z}_M \). We then approximate the missing modality using the \( i \)-th cluster center from the corresponding global cluster centers. The prediction for each instance is given by:
\begin{equation}
\hat{y}_c^i = \begin{cases}
g(\mathbf{z}_P, \mathbf{C}_{M,y}^{\text{global},(i)}) & \text{for PET-only instance}, \\
g(\mathbf{C}_{P,y}^{\text{global},(i)}, \mathbf{z}_M) & \text{for MRI-only instance},
\end{cases}
\end{equation}
where \( \mathbf{C}_{M,y}^{\text{global},(i)} \) and \( \mathbf{C}_{P,y}^{\text{global},(i)} \) represents the \( i \)-th cluster center from \( \mathbf{C}_{M,y}^{\text{global}} \) and \( \mathbf{C}_{P,y}^{\text{global}} \) respectively.

The loss for a PET-only instance is then computed as:
\begin{equation}
\mathcal{L}_{\text{mc}}(d_P) =  \frac{1}{T_{M,y}} \sum_{i=1}^{|\mathbf{S}_{M,y}^{\text{global}}|} \mathbf{S}_{M,y}^{\text{global},(i)} \cdot \mathcal{L}_{\text{CE}}(\hat{y}_c^i, y),
\end{equation}
where \( \mathcal{L}_{\text{CE}} \) denotes the cross-entropy loss function, and \( T_{M,y} = \sum_{i=1}^{|\mathbf{S}_{M,y}^{\text{global}}|} \mathbf{S}_{M,y}^{\text{global},(i)} \).

Similarly, for an MRI-only instance, the loss is computed as:
\begin{equation}
\mathcal{L}_{\text{mc}}(d_M) =  \frac{1}{T_{P,y}} \sum_{i=1}^{|\mathbf{S}_{P,y}^{\text{global}}|} \mathbf{S}_{P,y}^{\text{global},(i)} \cdot \mathcal{L}_{\text{CE}}(\hat{y}_c^i, y),
\end{equation}
where \( T_{P,y} = \sum_{i=1}^{|\mathbf{S}_{P,y}^{\text{global}}|} \mathbf{S}_{P,y}^{\text{global},(i)} \).

For any client $i$, let \( \mathcal{D}_i^{\text{single}} = \{ d \mid d \in \mathcal{D}_i, \, d \text{ is a single modality instance} \} \) be the subset of instances with a single modality (either PET-only or MRI-only). The overall modality completion loss is calculated as:
\begin{equation}
    \mathcal{L}_{\text{MC}}(\mathcal{D}_i)=\mathbb{E}_{d\in\mathcal{D}_{\text{single}}}\left[\mathcal{L}_{\text{mc}}(d)\right].
\end{equation}

In this manner, we use the global cluster centers of the available modality to approximate the missing modality, allowing the model to still make meaningful predictions in the presence of incomplete data.

\subsection{Overall Loss Function}

For client \( i \) with dataset \( \mathcal{D}_i \), the prediction \( \hat{y}_i \) is generated as described in \Cref{sec: preliminary} for each instance \( d_i \). Let \( \hat{\mathbf{y}} = (\hat{y}_1, \hat{y}_2, \dots, \hat{y}_{n_i}) \) denote the vector of predicted labels for the \( m \) instances in the client's dataset, and \( \mathbf{y} = (y_1, y_2, \dots, y_{n_i}) \) denote the corresponding true labels. The overall loss is computed as:

\begin{equation}
\mathcal{L}(\mathcal{D}_i) = \mathcal{L}_{\text{CE}}(\hat{\mathbf{y}}, \mathbf{y}) + \lambda_1 \mathcal{L}_{\text{CTR}}(\mathcal{D}_i) + \lambda_2 \mathcal{L}_{\text{MC}}(\mathcal{D}_i),
\end{equation}
where \( \lambda_1 \) and \( \lambda_2 \) are the regularization coefficients that balance the contributions of the contrastive loss and the modality completion loss, respectively.

\subsection{Modality-Aware Aggregation}

In this section, we describe the modality-aware aggregation (MAA) method adopted in our framework. Specifically, we use different aggregation weights for different modules within the model, based on the number of instances for each modality at each client.

Let \( n_P^i \) represent the number of instances with the PET modality at client \( i \), and \( n_M^i \) represent the number of instances with the MRI modality at client \( i \). The total numbers of instances with PET and MRI modalities across all clients are denoted as \( n^{\text{total}}_P \) and \( n^{\text{total}}_M \), and are computed as:
\begin{equation}
n^{\text{total}}_P = \sum_{i=1}^{N} n_P^i, \quad n^{\text{total}}_M = \sum_{i=1}^{N} n_M^i.    
\end{equation}

For client \( i \), the aggregation weights for the encoders are computed as:
\begin{equation}
w_P^i = \frac{n_P^i}{n_{\text{total}}^P}, \quad w_M^i = \frac{n_M^i}{n_{\text{total}}^M},
\end{equation}
where \( w_P^i \) and \( w_M^i \) are the aggregation weights for encoder $f_P$ and encoder $f_M$, respectively. 

The aggregation processes for encoders $f_P$ and $f_M$ are as follows:
\begin{equation}
    \mathbf{f}_P^{\text{global}} = \sum_{i=1}^{N} w_P^i \mathbf{f}_P^i, \quad
\mathbf{f}_M^{\text{global}} = \sum_{i=1}^{N} w_M^i \mathbf{f}_M^i,
\end{equation}
where \( \mathbf{f}_P^i \), \( \mathbf{f}_M^i \) are the model parameters of the PET encoder and MRI encoder, respectively.

For the classifier \( g \), the aggregation process is given by:
\begin{equation}
    \mathbf{g}^{\text{global}} = \frac{1}{\sum_{i=1}^{N}n_i}\sum_{i=1}^{N} n_i\cdot\mathbf{g}^i,
\end{equation}
where \( \mathbf{g}^i \) denotes the model parameters for classifier $g$.


               
\section{EXPERIMENT}
\subsection{Experimental Setup}
\begin{table*}[t]
\resizebox{\textwidth}{!}{
\begin{tabular}{cccccccccccc}
\hline
\multirow{2}{*}{$\alpha$} & \multirow{2}{*}{Method} & \multicolumn{5}{c}{$\beta = 0.2$}                                                                                                         & \multicolumn{5}{c}{$\beta = 0.4$}                                                                                     \\ \cline{3-12} 
                          &                         & Precision                 & Recall                    & F1                        & Accuracy                  & AUC                       & Precision             & Recall                & F1                    & Accuracy              & AUC                   \\ \hline
\multirow{7}{*}{0.2}      & FedAvg                  & 53.94 $\pm$ 3.56          & 54.05 $\pm$ 3.53          & 53.21 $\pm$ 3.90          & 53.41 $\pm$ 3.71          & 69.39 $\pm$ 2.05          & 53.58 $\pm$ 2.34      & 53.80 $\pm$ 2.77      & 52.30 $\pm$ 2.32      & 52.64 $\pm$ 2.14      & 68.20 $\pm$ 2.72      \\
                          & FedProx                 & 55.81 $\pm$ 1.58          & 54.37 $\pm$ 5.19          & 54.32 $\pm$ 2.21          & 54.62 $\pm$ 2.18          & 68.29 $\pm$ 2.93          & 54.39 $\pm$ 2.69      & 54.36 $\pm$ 1.82      & 53.27 $\pm$ 2.99      & 53.52 $\pm$ 2.97      & 68.11 $\pm$ 2.74      \\
                          & FedMed-GAN              & 54.62 $\pm$ 1.02          & 55.18 $\pm$ 2.73          & 53.46 $\pm$ 0.97          & 53.52 $\pm$ 1.07          & 68.80 $\pm$ 2.16          & 53.87 $\pm$ 3.66      & 54.66 $\pm$ 3.57      & 53.35 $\pm$ 3.61      & 53.41 $\pm$ 3.63      & 68.09 $\pm$ 2.53      \\
                          & FedMI                   & 54.77 $\pm$ 2.94          & 55.22 $\pm$ 4.36          & 54.26 $\pm$ 2.82          & 54.40 $\pm$ 2.83          & 69.15 $\pm$ 2.73          & 54.03 $\pm$ 3.88      & 54.46 $\pm$ 3.56      & 53.54 $\pm$ 4.22      & 53.63 $\pm$ 4.14      & 67.61 $\pm$ 3.93      \\
                          & MFCPL                   & 54.54 $\pm$ 1.88          & \textbf{56.06 $\pm$ 4.14} & 53.88 $\pm$ 1.64          & 54.07 $\pm$ 1.93          & 69.39 $\pm$ 2.71          & 53.86 $\pm$ 5.42      & 54.01 $\pm$ 5.44      & 53.22 $\pm$ 5.16      & 53.30 $\pm$ 5.34      & 67.58 $\pm$ 3.67      \\
                          & PmcmFL                  & 53.17 $\pm$ 1.60          & 52.73 $\pm$ 3.50          & 52.58 $\pm$ 1.81          & 52.64 $\pm$ 2.03          & 67.51 $\pm$ 4.07          & 49.03 $\pm$ 1.55      & 48.01 $\pm$ 3.03      & 48.38 $\pm$ 2.00      & 48.46 $\pm$ 2.31      & 63.71 $\pm$ 2.50      \\
                          & ClusMFL                    & \textbf{57.16 $\pm$ 2.32}     & 54.73 $\pm$ 3.93              & \textbf{56.56 $\pm$ 2.36}     & \textbf{56.92 $\pm$ 2.41}     & \textbf{72.81 $\pm$ 3.64}     & \textbf{56.06 $\pm$ 1.31} & \textbf{55.44 $\pm$ 2.19} & \textbf{55.38 $\pm$ 1.07} & \textbf{55.49 $\pm$ 1.10} & \textbf{72.50 $\pm$ 2.02} \\ \hline
\multirow{7}{*}{0.4}      & FedAvg                  & 53.75 $\pm$ 3.86          & 54.26 $\pm$ 3.76          & 52.76 $\pm$ 4.64          & 53.08 $\pm$ 4.34          & 67.91 $\pm$ 3.48          & 52.60 $\pm$ 3.00      & 53.95 $\pm$ 3.66      & 51.84 $\pm$ 3.06      & 51.98 $\pm$ 3.07      & 67.03 $\pm$ 3.22      \\
                          & FedProx                 & 54.36 $\pm$ 3.50          & 54.28 $\pm$ 4.17          & 53.77 $\pm$ 3.48          & 53.96 $\pm$ 3.51          & 68.26 $\pm$ 3.53          & 52.16 $\pm$ 2.20      & 52.93 $\pm$ 3.76      & 51.35 $\pm$ 2.13      & 51.54 $\pm$ 2.07      & 67.43 $\pm$ 3.64      \\
                          & FedMed-GAN              & 52.97 $\pm$ 1.63          & 52.89 $\pm$ 2.41          & 52.43 $\pm$ 1.74          & 52.53 $\pm$ 1.63          & 67.25 $\pm$ 3.45          & 53.56 $\pm$ 3.57      & 54.02 $\pm$ 3.95      & 52.54 $\pm$ 3.97      & 52.86 $\pm$ 3.63      & 67.91 $\pm$ 2.93      \\
                          & FedMI                   & 55.40 $\pm$ 2.85          & 53.82 $\pm$ 2.45          & 54.84 $\pm$ 2.75          & 54.95 $\pm$ 2.69          & 68.59 $\pm$ 2.48          & 54.01 $\pm$ 2.07      & 52.36 $\pm$ 1.98      & 53.62 $\pm$ 2.14      & 53.85 $\pm$ 2.06      & 66.70 $\pm$ 2.63      \\
                          & MFCPL                   & 55.07 $\pm$ 4.34          & 54.66 $\pm$ 3.93          & 54.34 $\pm$ 4.37          & 54.51 $\pm$ 4.57          & 66.60 $\pm$ 2.58          & 52.84 $\pm$ 4.14      & 54.69 $\pm$ 3.92      & 51.70 $\pm$ 3.89      & 51.87 $\pm$ 3.74      & 67.81 $\pm$ 3.48      \\
                          & PmcmFL                  & 51.71 $\pm$ 5.54          & 49.80 $\pm$ 6.21          & 50.77 $\pm$ 5.21          & 50.88 $\pm$ 5.42          & 65.54 $\pm$ 6.22          & 52.45 $\pm$ 3.96      & 50.03 $\pm$ 5.74      & 50.40 $\pm$ 4.72      & 50.77 $\pm$ 4.66      & 63.31 $\pm$ 5.17      \\
                          & ClusMFL                    & \textbf{56.59 $\pm$ 4.53} & \textbf{54.68 $\pm$ 4.49} & \textbf{56.17 $\pm$ 4.08} & \textbf{56.37 $\pm$ 4.10} & \textbf{73.25 $\pm$ 4.11} & \textbf{54.83 $\pm$ 6.13} & \textbf{54.88 $\pm$ 6.09} & \textbf{54.22 $\pm$ 5.89} & \textbf{54.40 $\pm$ 6.03} & \textbf{72.22 $\pm$ 4.47} \\ \hline
\end{tabular}
}
\caption{Performance Comparison Across Different Federated Learning Methods (Mean $\pm$ Standard Deviation \%) under Different Settings.}
\label{main result}
\end{table*}

\begin{figure*}[h]
    \centering
    \includegraphics[width=\linewidth]{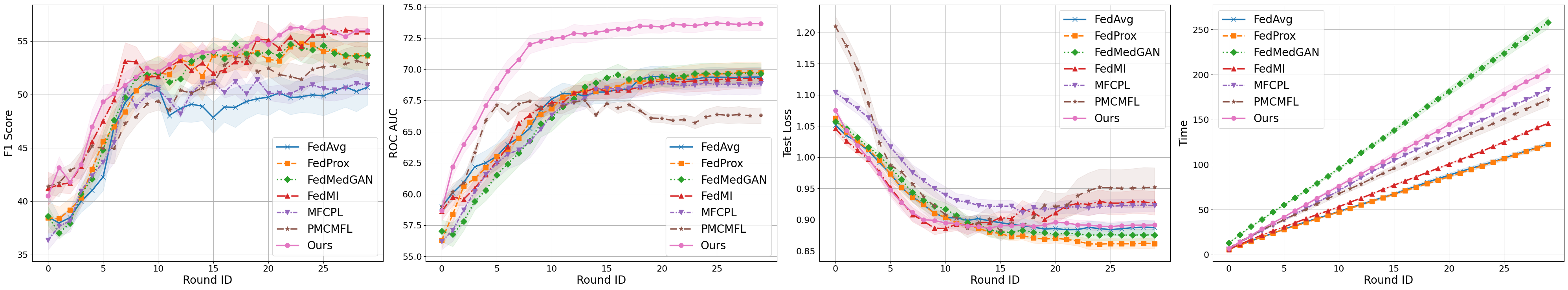}
    \caption{Training curves of different methods.}
    \label{fig:training curve}
\end{figure*}
The brain imaging dataset used in this study is sourced from the Alzheimer's Disease Neuroimaging Initiative (ADNI) public repository~\cite{mueller2005alzheimer} and comprises 915 participants stratified into three diagnostic categories: 297 healthy controls (HC), 451 mild cognitive impairment (MCI), and 167 Alzheimer's disease (AD) patients. Multimodal neuroimaging acquisitions encompass structural Magnetic Resonance Imaging (VBM-MRI) and 18 F-florbetapir PET (AV45-PET), enabling examination of brain structure and amyloid plaque deposition, respectively.

Consistent with established neuroimaging processing pipelines \cite{barshan2015stage,zhu2010graphene}, we preprocess neuroimaging data to region-of-interest (ROI) features from each participant’s images. 
First, the multi-modality imaging scans are aligned to each participant's same visit. All imaging scans are aligned to a T1-weighted template image. 
Subsequently, the images are segmented into gray matter (GM), white matter (WM) and cerebrospinal fluid (CSF) maps. 
They are normalized to the standard Montreal Neurological Institute (MNI) space as $2 \times 2 \times 2$ mm$^3$ voxels, being smoothed with an $8$ mm full-width at half-maximum (FWHM) Gaussian kernel.
We preprocess the structural MRI scans with voxel-based morphometry (VBM) by using the SPM software \cite{ashburner2000voxel}, and register the AV45-PET scans to the MNI space by SPM. 
For both MRI and PET scans, we parcellate the brain into 90 ROIs (excluding the cerebellum and vermis) based on the AAL-90 atlas \cite{tzourio2002automated}, and computed ROI-level measures by averaging voxel-wise values within each region.

In the original dataset, each instance contains both modalities. We first split the dataset into a training set and a test set with a ratio of 1:4. For the test set, we ensure that the proportions of the three types of instances are equal, \textit{i.e.}, each type accounts for $\frac{1}{3}$ of the total. 
In the federated learning setup, we distribute the training instances equally across clients while preserving label distribution. Each instance is then assigned a type (multimodal or single-modality) based on the MFL settings controlled by $\alpha_1$, $\alpha_2$, $\beta_1$, and $\beta_2$ as described in \Cref{setting}. Single-modality instances drop the corresponding modality. For simplicity, we set $\alpha_1 = \alpha_2 = \alpha$ and $\beta_1 = \beta_2 = \beta$.

To evaluate the effectiveness of our proposed method, we compare it against several baseline approaches, including FedAvg \cite{FedAvg}, FedProx \cite{FedProx}, FedMed-GAN \cite{FedMed-GAN}, FedMI \cite{FedMI}, MFCPL \cite{MFCPL}, and PmcmFL \cite{PmcmFL}. FedAvg and FedProx are traditional federated learning algorithms, with FedAvg employing simple parameter averaging, while FedProx introduces a proximal term to the objective to mitigate client heterogeneity. FedMed-GAN employs CycleGAN \cite{cyclegan} to complete missing modalities, enhancing diagnosis accuracy. FedMI, MFCPL, and PmcmFL leverage prototype learning to model class distributions and align features, effectively addressing incomplete modalities and heterogeneous data across clients. All methods are evaluated under the same experimental setup to ensure a fair comparison.

In our experiment, we set the number of clients to $N=10$, with the number of communication rounds fixed at 30 and each client performing 10 local training epochs. For optimization, we employ the Adam \cite{adam} optimizer with an initial learning rate of 0.01. To dynamically adjust the learning rate during training, we use cosine annealing strategy. We conduct 5-fold cross-validation and report the results as the mean $\pm$ standard deviation across all folds. For evaluation metrics, we adopt a weighted average for precision, F1-score, and ROC-AUC, while using a macro average for recall.

\subsection{Main Results}
\Cref{main result} presents a comprehensive comparison of the performance of our proposed method against several baseline approaches. The experiments were conducted under varying configurations of $\beta$ and $\alpha$, which denote modality incompleteness and client diversity, respectively.

As indicated in \Cref{main result}, our method consistently outperforms the baseline approaches across different settings of modality incompleteness, achieving superior results in terms of precision, recall, F1 score, accuracy, and AUC. Notably, certain algorithms specifically designed for modality-incomplete MFL fail to outperform traditional federated learning methods, such as FedProx, in some scenarios.

Moreover, as the values of $\alpha$ and $\beta$ increase—corresponding to a higher proportion of instances with only a single modality—most of the baseline algorithms exhibit a noticeable decline in performance. In contrast, our proposed method demonstrates stable and robust performance, highlighting its effectiveness and resilience in handling varying degrees of modality incompleteness.

\subsection{Ablation Study}

 In order to assess the contributions of each component, we conduct an ablation study under the setting of $\alpha = 0.4$ and $\beta = 0.2$, as shown in Table \ref{table:ablation study}. The results demonstrate that applying modality-aware aggregation (MAA) alone yields lower performance across all metrics. Incorporating the contrastive loss ($\mathcal{L}_{\text{CTR}}$) improves the results significantly, while the modality completion loss ($\mathcal{L}_{\text{MC}}$) also leads to moderate gains. Combining both losses without MAA further enhances performance. The full method, which includes both MAA and the two loss functions, achieves the highest performance, with precision, recall, F1-score, accuracy, and AUC all showing notable improvements. These findings highlight the effectiveness of combining modality-aware aggregation with the contrastive and modality completion losses in addressing modality incompleteness in MFL.
\begin{table}[]
\resizebox{\columnwidth}{!}{%
\begin{tabular}{cccccccc}
\hline
MAA          & $\mathcal{L}_{\text{CTR}}$ & $\mathcal{L}_{\text{MC}}$ & Precision    & Recall       & F1           & Accuracy          & AUC          \\ \hline
$\checkmark$ &                            &                           & 53.74 $\pm$ 5.81 & 53.62 $\pm$ 7.24 & 52.66 $\pm$ 5.45 & 52.86 $\pm$ 5.54 & 72.10 $\pm$ 5.10 \\
             & $\checkmark$               &                           & 54.89 $\pm$ 6.49 & 54.12 $\pm$ 6.42 & 53.99 $\pm$ 6.23 & 54.18 $\pm$ 6.39 & 72.42 $\pm$ 4.22 \\
             &                            & $\checkmark$              & 53.50 $\pm$ 1.60 & 54.08 $\pm$ 1.80 & 53.02 $\pm$ 1.85 & 53.19 $\pm$ 1.80 & 71.34 $\pm$ 2.69 \\ \hline
             & $\checkmark$               & $\checkmark$              & 56.39 $\pm$ 2.95 & 53.69 $\pm$ 3.39 & 55.47 $\pm$ 2.58 & 55.82 $\pm$ 2.84 & 71.29 $\pm$ 3.35 \\
$\checkmark$ &                            & $\checkmark$              & 54.67 $\pm$ 2.21 & 52.27 $\pm$ 2.50 & 54.13 $\pm$ 2.16 & 54.40 $\pm$ 2.27 & 71.41 $\pm$ 2.60 \\
$\checkmark$ & $\checkmark$               &                           & 55.40 $\pm$ 0.78 & 53.48 $\pm$ 2.33 & 54.94 $\pm$ 0.91 & 55.16 $\pm$ 1.00 & 71.61 $\pm$ 2.38 \\ \hline
$\checkmark$ & $\checkmark$               & $\checkmark$              &\textbf{56.59 $\pm$ 4.53} & \textbf{54.68 $\pm$ 4.49} & \textbf{56.17 $\pm$ 4.08} & \textbf{56.37 $\pm$ 4.10}  &   \textbf{73.25 $\pm$ 4.11}      \\ \hline
\end{tabular}
}
\caption{Ablation Study: Performance Evaluation under the Setting of $\alpha = 0.4$, $\beta = 0.2$ (Mean $\pm$ Standard Deviation \%)}
\label{table:ablation study}
\vspace{-25pt}
\end{table}
\subsection{Convergence Efficiency}

To better analyze the convergence behavior of different algorithms in modality-incompleteness scenarios, we conduct experiments with four different random seeds in the setting of the first fold, with $\alpha = 0.4$ and $\beta = 0.2$. The training curves, displayed in \Cref{fig:training curve}, represent the mean values across these random seeds, while the shaded areas indicate the standard deviation. From \Cref{fig:training curve}, we make the following observations:

\begin{itemize}
    \item \textbf{F1 Score and AUC:} Our method exhibits a rapid and consistent increase in both F1 score and AUC during the initial communication rounds, and subsequently stabilizes at higher values compared to the baseline methods. This indicates its superior ability to quickly capture relevant patterns. Furthermore, the consistent improvement in both metrics suggests that our approach enhances class discrimination more effectively as training progresses.
    \item \textbf{Test Loss:} Our method achieves convergence with fewer communication rounds compared to the baseline methods. Specifically, it converges around the 11th communication round, whereas most baseline methods require approximately 20 rounds to reach convergence, highlighting the superior efficiency of our approach.
    \item \textbf{Time:} Our method achieves significant performance improvements without a substantial increase in training time. Furthermore, the training time of our method remains significantly lower than that of FedMed-GAN, which requires adversarial training for GANs.
\end{itemize}

\section{CONCLUSION}
In this study, we propose ClusMFL, a novel framework designed to address modality incompleteness in multimodal federated learning. ClusMFL enhances local data representation with the FINCH clustering algorithm, mitigates missing modality effects through supervised contrastive loss and modality completion loss, and employs a modality-aware aggregation strategy to adaptively integrate client contributions. Extensive experiments on the ADNI dataset demonstrate that ClusMFL outperforms state-of-the-art methods, particularly in scenarios with severe modality incompleteness, though at a slightly higher computational cost. Future research will explore the incorporation of additional modalities to further enhance the proposed framework's applicability.
\addtolength{\textheight}{-12cm}   









\bibliographystyle{IEEEtran}

\end{document}